\documentclass[a4paper,aps,prx,reprint,superscriptaddress,floatfix,longbibliography]{revtex4-2}
\usepackage{dcolumn}	
\usepackage{graphicx}
\usepackage[usenames,dvipsnames]{xcolor}
\usepackage{amsmath}
\usepackage{microtype}
\usepackage{times}
\usepackage{changes}
\usepackage{soul}
\usepackage{hyperref}
\hypersetup{
  colorlinks,
  allcolors=NavyBlue,
  linktoc=all
}

\usepackage{siunitx}

\usepackage{soul}
\usepackage[normalem]{ulem}

\newcommand*\dagg{^{\dagger}}
\newcommand*\sstar{^{*}}
\newcommand*{\s}[1]{\ensuremath{_\mathrm{#1}}}	

\begin{document}

\title{Dissipative Quantum Feedback in Measurements Using a Parametrically Coupled Microcavity
}
\author{Liu Qiu}
\altaffiliation{Present address: Institute of Science and Technology Austria, Am Campus 1, 3400 Klosterneuburg, Austria}
\email{liu.qiu@ist.ac.at}
\author{Guanhao Huang}
\affiliation{Institute of Physics, Swiss Federal Institute of Technology Lausanne (EPFL), CH-1015 Lausanne, Switzerland}
\author{Itay Shomroni}
\affiliation{Institute of Physics, Swiss Federal Institute of Technology Lausanne (EPFL), CH-1015 Lausanne, Switzerland}
\affiliation{The Racah Institute of Physics, The Hebrew University of Jerusalem, Jerusalem 9190401, Israel}
\author{Jiahe Pan}
\affiliation{Institute of Physics, Swiss Federal Institute of Technology Lausanne (EPFL), CH-1015 Lausanne, Switzerland}
\author{Paul Seidler}
\affiliation{IBM Quantum, IBM Research Europe, Zurich, S\"aumerstrasse 4, CH-8803 R\"uschlikon, Switzerland}
\author{Tobias J. Kippenberg}
\email{tobias.kippenberg@epfl.ch}
\affiliation{Institute of Physics, Swiss Federal Institute of Technology Lausanne (EPFL), CH-1015 Lausanne, Switzerland}

\date{\today}

\begin{abstract}
Micro- and nanoscale optical or microwave cavities are used in a wide range of classical applications and quantum science experiments, ranging from precision measurements, laser technologies to quantum control of mechanical motion. The dissipative photon loss via absorption, present to some extent in any optical cavity, is known to introduce thermo-optical effects and thereby impose fundamental limits on precision measurements. Here, we theoretically and experimentally reveal that such dissipative photon absorption can result in quantum feedback via in-loop field detection of the absorbed optical field, leading to the intracavity field fluctuations to be squashed or antisquashed. A closed-loop dissipative quantum feedback to the cavity field arises. Strikingly, this modifies the optical cavity susceptibility in coherent response measurements (capable of both increasing or decreasing the bare cavity linewidth) and causes excess noise and correlations in incoherent interferometric optomechanical measurements using a cavity, that is parametrically coupled to a mechanical oscillator. We experimentally observe such unanticipated dissipative dynamics in optomechanical spectroscopy of sideband-cooled optomechanical crystal cavitiess at both cryogenic temperature (approximately 8 K) and ambient conditions. The dissipative feedback introduces effective modifications to the optical cavity linewidth and the optomechanical scattering rate and gives rise to excess imprecision noise in the interferometric quantum measurement of mechanical motion. Such dissipative feedback differs fundamentally from a quantum nondemolition feedback, e.g., optical Kerr squeezing. The dissipative feedback itself always results in an antisqueezed out-of-loop optical field, while it can enhance the coexisting Kerr squeezing under certain conditions. Our result applies to cavity spectroscopy in both optical and superconducting microwave cavities, and equally applies to any dissipative feedback mechanism of different bandwidth inside the cavity. It has wide-ranging implications for future dissipation engineering, such as dissipation enhanced sideband cooling and Kerr squeezing, quantum frequency conversion, and nonreciprocity in photonic systems.
\end{abstract}

\pacs{Valid PACS appear here}

\maketitle
Dissipation is the hallmark of an open quantum system, which leads to undesired decoherence and hinders the observation of quantum phenomenon.
The developments in nanofabrication  over the last decades  enable engineered low-loss optical micro-cavities for a wide range of novel physics and applications in cavity specroscopy~\cite{vahala_optical_2003,gaeta_photonicchipbased_2019,lodahl_interfacing_2015,aspelmeyer_cavity_2014}, from  frequency metrology, to cavity quantum electrodynamics and cavity optomechanics. 
Despite these advancements~\cite{akahane_high_2003,armani_ultrahigh_2003,burek_high_2014,liu_highyield_2020}, dissipative absorption of photons is ubiquitous in optical cavities of different scales. 
Such photon absorption manifests as a feedback mechanism to the optical cavity field, e.g. by changing cavity properties. 
Optical absorption is widely employed in photonic technologies~\cite{carmon_dynamical_2004a,almeida_optical_2004,nozaki_subfemtojoule_2010,favero_spheroidal_2012,heylman_optical_2016},
 such as thermal tuning, optical switching, and sensing. 
The incoherent photon absorption has been shown theoretically and experimentally to limit precision measurements~\cite{braginsky_thermodynamical_1999,goda_photothermal_2005,matsko_whisperinggallerymode_2007,sun_squeezing_2017,drake_thermal_2020}, such as in the gravitational wave detection~\cite{braginsky_thermodynamical_1999}, position measurements~\cite{deliberato_quantum_2011,restrepo_classical_2011}, and frequency metrology~\cite{stone_thermal_2018,drake_thermal_2020}. 
However, the dissipative dynamics due to photon absorption induced feedback in 
the widely used micro-cavity spectroscopy still remains unexplored.

We show that a closed-loop dissipative quantum  feedback arises in the micro-cavity spectroscopy due to photon absorption~\cite{shapiro_theory_1987,taubman_intensity_1995a,wiseman_squashed_1999}.
The optical absorption can be modeled by an in-loop detection of the absorbed photons, whose fluctuations can be squashed or anti-squashed. 
The stochastic fluctuations in absorption lead to cavity frequency fluctuations,
which results in a feedback loop to the cavity field~\cite{taubman_intensity_1995a,wiseman_inloop_1998,wiseman_squashed_1999}.
The dissipative feedback gives rise to a modified optical cavity susceptibility,
and reshapes the optical transduction for the probing field and intracavity field.
The optical cavity under such condition is equivalently coupled to a dissipative reservoir, whose noise properties are different from the original vacuum noise, which causes excess noise in the out-of-loop measured optical field.
The dissipative feedback introduces excess classical correlations in interferometric quantum measurement due to uncorrelated noise from the in-loop detection, and masks the micro-cavity spectroscopy, similar to laser noise~\cite{kippenberg_phase_2013,rabl_phasenoise_2009,safavi-naeini_laser_2013}.

More specifically, we report the observation of such dissipative dynamics in optomechanical spectroscopy in sideband cooling of a 5GHz breathing mode in optomechanical crystal cavities (OMC)~\cite{eichenfield_optomechanical_2009a,chan_optimized_2012,qiu_floquet_2019,qiu_laser_2020,shomroni_optical_2019,shomroni_twotone_2019}.
The optomechanical interaction via radiation pressure in the cavity has received significant interests over the last decade~\cite{aspelmeyer_cavity_2014},
from ground state cooling~\cite{wilson-rae_theory_2007,marquardt_quantum_2007,chan_laser_2011,teufel_sideband_2011,wilson_measurementbased_2015,rossi_measurementbased_2018,qiu_laser_2020},
quantum correlations~\cite{aggarwal_roomtemperature_2020,kampel_improving_2017,purdy_strong_2014,safavi-naeini_squeezed_2013,sudhir_quantum_2017,safavi-naeini_observation_2012,weinstein_observation_2014,sudhir_appearance_2017},
to a variety of recently emerged applications, such as nonreciprocal devices~\cite{bernier_nonreciprocal_2017,peterson_demonstration_2017,xu_nonreciprocal_2019}, and quantum transducers~\cite{hill_coherent_2012,arnold_converting_2020,bochmann_nanomechanical_2013,higginbotham_harnessing_2018,jiang_efficient_2020,mirhosseini_superconducting_2020,wu_microwavetooptical_2020}.
Macroscopic quantum effects have been explored~\cite{hong_hanbury_2017,marinkoviifmmodeacutecelsecfi_optomechanical_2018,ockeloen-korppi_stabilized_2018,riedinger_remote_2018,safavi-naeini_observation_2012,shomroni_optomechanical_2020,weinstein_observation_2014,kronwald_arbitrarily_2013,lecocq_quantum_2015,pirkkalainen_squeezing_2015,wollman_quantum_2015},
including motional sideband asymmetry~\cite{safavi-naeini_observation_2012,weinstein_observation_2014,purdy_optomechanical_2015,underwood_measurement_2015,sudhir_appearance_2017,qiu_floquet_2019}, mechanical entanglement~\cite{ockeloen-korppi_stabilized_2018,riedinger_remote_2018} and squeezing~\cite{kronwald_arbitrarily_2013,lecocq_quantum_2015,pirkkalainen_squeezing_2015,wollman_quantum_2015}.
However, most of these quantum optomechanical protocols are currently only feasible in very few optomechanical systems~\cite{chan_laser_2011,peterson_laser_2016,teufel_sideband_2011,tsaturyan_ultracoherent_2017,verhagen_quantumcoherent_2012}, mainly due to the ubiquitous dissipative photon absorption.
More specifically, we perform optomechanical spectroscopy with two OMCs of different optical absorption losses  at cryogenic temperature ($\sim$ 8K) and under ambient conditions respectively.
We observe a modified and power dependent optical cavity linewidth in the coherent cavity response, due to the dissipative feedback.
At cryogenic temperature ($\sim$ 8K), an anti-squashed in-loop absorbed optical field leads to an effective narrowing of the optical cavity linewidth,
which may be exploited for enhanced sideband cooling and even normal mode splitting~\cite{rossi_enhancing_2017a,rossi_normalmode_2018,zippilli_cavity_2018}.
The incoherent optomechanical scattering rate is modified accordingly due to the effective cavity linewidth, profoundly influencing mechanical quantum measurements.
The dissipative feedback also results in \emph{excess noise} in the quantum measurement of the mechanical motion in the balanced heterodyne detection.

We compare such dissipative dynamics to a quantum non-demolition (QND)
feedback~\cite{shapiro_theory_1987,haus_theory_1986a}, i.e. optical Kerr squeezing, in on-chip micro-cavities.
The dissipative photon absorption itself always generates excess noise in the out-of-loop optical field,
regardless of the fluctuations in the in-loop absorbed optical field~\cite{taubman_intensity_1995a,wiseman_inloop_1998,wiseman_squashed_1999}.
We identify a regime where the dissipative absorption can even improve the coexisting Kerr squeezing,
which can be attributed to the coherent dissipative dynamics, and is evaluated for a state-of-art $\textrm{Si\s{3}N\s{4}}$  micro-resonator~\cite{liu_highyield_2020}.
\section{Theory}
We first consider an optical cavity of resonant frequency $\omega_c$, which is coupled to an input field $\hat{a}\s{ex,in} = (\bar{a}\s{in}+\delta a\s{in}) e^{-i\omega_L t}$ of frequency $\omega_L$ at a rate of $\kappa\s{ex}$.
In the rotating frame of the pumping frequency $\omega_L$,
the dynamics of the intracavity field $\hat{a}$ can be obtained through the quantum Langevin equation, considering the dissipative photon absorption,
\begin{equation}
	\dot{\hat a} =  [i\left(\Delta-\Delta\s{th}\right)-\frac{\kappa}{2}]\hat a
	 +\sqrt{\kappa{\s{ex}}} \hat a \s{ex,in}
	+\sqrt{\kappa{\s{s}}}  \delta \hat a \s{s,in}
	+\sqrt{\kappa{\s{a}}}  \delta \hat a \s{a,in}
\label{eq:qle}
\end{equation}
with $\Delta = \omega_L-\omega_c$, and $\Delta{\s{th}}=g\s{th} \delta T$ the cavity frequency shift via optical absorption induced temperature change $\delta T$ of coefficient $g\s{th}$. 
The intrinsic cavity losses $\kappa_0$ in this treatment is separated into an absorption loss channel with rate $\kappa\s{a}$ (the equivalent rate of photons absorbed which lead to local heating of the cavity material, followed by a photo-thermal induced cavity frequency shift) as well as a scattering loss channel with rate $\kappa_{s}$ (which does not lead to a local heating of the cavity material).
As a result of the presence of two dissipation channels, from the input-output relations for an open quantum system~\cite{gardiner_input_1985}, the optical cavity is driven by input vacuum fluctuations $\delta \hat{a}\s{a,in}$ and $\delta \hat{a}\s{s,in}$ from both channels (cf. Eq.(\ref{eq:qle}) right hand side), with a total cavity loss being $\kappa=\kappa\s{ex}+\kappa_0$.
In the weak coupling regime, we can linearize the intracavity field $\hat{a}(t) = \bar {a}(t) + \delta \hat{a}(t)$, with $\bar{a}$ the mean field amplitude and $\delta \hat{a}$ the field fluctuation.

We adopt a quantum treatment of the dissipative photon absorption, whose field amplitude can be obtained by, 
\begin{equation}
\hat{a}\s{a,out} = \delta \hat{a}\s{a,in}-\sqrt{\kappa\s{a}}\hat{a},
\end{equation}
with corresponding absorbed photon flux,
\begin{equation}
\hat{I}\s{a} = \hat{a}^{\dagger}\s{a,out}  \hat{a}\s{a,out}.
\end{equation}
The absorption photon flux is composed of two terms, i.e. a DC term $\bar{I}\s{a}=\kappa\s{a} |\bar{a}|^2$,
and a fluctuating term,
\begin{equation}
	\delta \hat{I}\s{a}=\kappa\s{a} (\bar{a} \delta \hat{a}\dagg + \bar{a}\sstar \delta \hat{a})-\sqrt{\kappa\s{a}} (\bar{a}\sstar \delta \hat{a}\s{a,in}+ \delta \hat{a}\dagg \s{a,in} \bar{a}).
\end{equation}

The dynamics of the temperature change due to the optical absorption is given by,
$\delta\dot T(t)=-\gamma\s{th}\delta T(t)+g_a \hat{I}\s{a}$,
where $g_a \hat{I}\s{a}$ is the rate of the temperature change due to the optical absorption and $\gamma\s{th}$ is the thermal decay rate.
The macroscopic aparatus, i.e.  $\delta T$, is not quantized, which is however correlated to the fluctuations of the in-loop photodetection represented with quantum operators.
For simplicity,  we assume a one pole model of thermal heating response. More details considering the geometric dependence are discussed in Appendix~\ref{app:simulation}.
The quantum Langevin equations of the linearized optical field fluctuations can be obtained in the frequency domain,
\begin{equation}
\begin{aligned}
	\chi^{-1}\s{c,0}(\Omega)\delta {\hat a}  =&
\sigma_{d}(\Omega) \kappa_{a}
(\delta \hat{a}+\delta \hat{a}\dagg)
 +\sqrt{\kappa \s{a}}\delta\hat a_d\\
 &+\sqrt{\kappa\s{ex}}\delta\hat a\s{ex,in}
 +\sqrt{\kappa_s}\delta\hat a\s{s,in},
 \end{aligned}
\label{eq:qlept}
\end{equation}
with the  \textit{intrinsic} optical susceptibility,
\begin{equation}
\chi\s{c,0}(\Omega)= 1/(\kappa/2 - i(\Omega+ \bar{\Delta})),
\label{eq:chic0}
\end{equation}
and $\bar{\Delta}= \Delta - g_a \kappa_a g\s{th}\bar{n}_c/\gamma\s{th}$,
incorporating the additional static cavity frequency shift.
The mean intracavity photon number is
$\bar{n}_c = |\bar{a}|^2 =\kappa\s{ex}|\bar{a}\s{in}|^2/(\kappa^2/4+\bar{\Delta}^2)$.
We note that, the intracavity field is coupled to an effective dissipative reservoir,
\begin{equation}
\delta \hat{a}_{d} =  \delta \hat{a}\s{a,in} -  \sigma_{d}(\Omega)\left( \delta \hat{a}\s{a,in} + \delta \hat{a}\s{a,in}\dagg \right),
\end{equation}
by coupling additionally to the amplitude quadrature $\left( \delta \hat{a}\s{a,in} + \delta \hat{a}\s{a,in}\dagg \right)$
 of the absorption field. Here we introduce a unitless photon-number-enhanced dissipation coefficient,
\begin{equation}
\sigma_d(\Omega) = \frac{ g_a  g\s{th} }{\Omega+ i \gamma\s{th}} \bar{n}_c,
\end{equation}
with $\sigma^{*}_d(-\Omega)=-\sigma_d(\Omega)$.
A constant single photon dissipation coefficient is given by $\sigma\s{0}(\Omega)=\sigma_d(\Omega)/\bar{n}_c$.

\begin{figure}
\centering
    \includegraphics[scale=1]{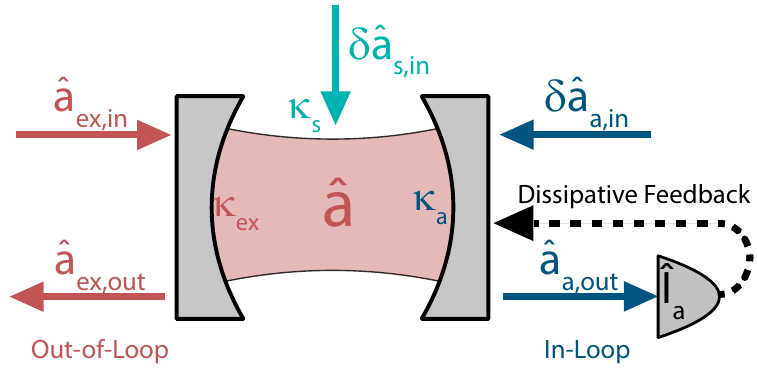}
  \caption{
  \textbf{Scheme for dissipative feedback in an optical micro-cavity. }
  The intracavity field $\hat{a}$ is coupled to several different reservoirs, i.e. the laser input $\hat{a}\s{ex,in}$, the vacuum noise  from the intrinsic optical loss $\delta \hat{a}\s{s,in}$ and the dissipative absorption $\delta \hat{a}\s{a,in}$. The in-loop measurement of absorption field $\hat{a}\s{a,out}$, results in the absorption flux $\hat{I}\s{a}$, which is fed back to the optical cavity (black dashed curve). The out-of-loop measurement of $\hat{a}\s{ex,out}$ is performed for coherent and incoherent spectroscopy.
}
\label{fig:scheme}
\end{figure}

Novel dynamics arises in the cavity field (cf. Eq.~\ref{eq:qlept}), 
which can be interpreted as a closed-loop dissipative feedback to optical cavity as shown in Fig.~\ref{fig:scheme}. 
The photon absorption manifests as an in-loop photodetection, which results in a feedback path to the optical cavity, i.e. by changing the  cavity frequency via photothermal effects.
The noise spectral density of the in-loop photon flux fluctuation 
$\delta I\s{a}$, when normalized to shot noise, is given by 
\begin{equation}
	S\s{\delta I_a}(\Omega)= \left|1-\chi\s{fb}(\Omega)\right|^{-2},
\end{equation}
where 
\begin{equation}
	\chi\s{fb} (\Omega)= \sigma_d(\Omega) \kappa_a \left(\chi\s{c,0}(\Omega)-\chi^{*}\s{c,0}(-\Omega)\right).
\end{equation}
The dissipative feedback results in squashed or anti-squashed in-loop optical field.
We can obtain the \emph{modified optical susceptibility}
\begin{equation}
	\chi\s{c,eff}(\Omega) =  \chi\s{c,0}(\Omega)/(1-\chi\s{fb}(\Omega)).
\end{equation}

In the case of  $\bar{\Delta}\ll-\kappa$ 
,  the PSD of the in-loop photon flux at $\Omega\sim|\bar{\Delta}|$ is,
\begin{equation}
	S\s{\delta I_a}(\Omega)\simeq\left|1-\frac{2\bar{n}_c \kappa_a g_a g\s{th}}{\Omega \kappa}\right|^{-2},
\end{equation}
and the effective optical susceptibility can be further simplified,
\begin{equation}
	\chi\s{c,eff}(\Omega) = \frac{1}{\kappa\s{eff}/2-i(\bar{\Delta}\s{eff}+\Omega)},
	\label{eq:chiceff}
\end{equation}
with a modified effective cavity linewidth
\begin{equation}
	\kappa\s{eff}= \kappa-2\kappa_a \operatorname{Re}(\sigma_d(\Omega)),
\label{eq:keff}
\end{equation}
and a modified effective detuning 
\begin{equation}
	\bar{\Delta}\s{eff}=  \bar{\Delta} + \kappa_a\operatorname{Im}(\sigma_d(\Omega)), 
\end{equation}
where
$\bar{\Delta}=\Delta - \bar{n}_c g_a\kappa_a g\s{th}/\gamma\s{th} $.
Due to the squashed or anti-squashed absorbed photon fluctuations, the dissipative feedback leads to modified cavity susceptibility for the input probing field.
In practice, the photothermal coefficient $g\s{th}$ can have different signs at room temperatures and cryogenic temperatures for different materials.
For $g\s{th}<0$, we have $\kappa\s{eff}   > \kappa$
and       $\bar{\Delta}\s{eff} > \bar{\Delta}$; while
for $g\s{th}>0$, we have $\kappa\s{eff}  < \kappa$
and       $\bar{\Delta}\s{eff} < \bar{\Delta}$.

As shown in Eq.~\ref{eq:qlept}, the intracavity field is coupled to an effective dissipative reservoir $\delta \hat{a}_d$.
Different from the vacuum noise, the noise operator for the dissipative reservoir $\delta\hat{a}_d$ satisfies the following correlations,
\begin{equation}
	\begin{aligned}
		\left<\delta \hat{a}_d\dagg(\Omega) \delta \hat{a}_d(\Omega^\prime)\right>=&
		-\sigma_d (\Omega)  \sigma_d (\Omega^\prime)
		\delta(\Omega+\Omega^\prime) 2\pi\\
		\left<\delta \hat{a}_d(\Omega) \delta \hat{a}\dagg_d(\Omega^\prime)\right>=& (1-\sigma_d (\Omega))(1+\sigma_d (\Omega^\prime))\delta(\Omega+\Omega^\prime) 2\pi,
	\end{aligned}
\end{equation}
which may result in incoherent excess noise and noise correlation for the intracavity field.

To give an example, we show how the dissipative feedback can result in modified cavity susceptibility and excess noise correlation in the quantum measurement of mechanical motion.
In such cavity optomechanical system, a mechanical mode of frequency $\Omega_m$ is dispersively coupled to an optical mode of frequency $\omega_c$ at a vacuum optomechanical coupling rate of $g_0$ via radiation pressure.
A cooling tone is applied close to the cavity red sideband, i.e. $\bar{\Delta}\simeq-\Omega_m$.
The cooling tone gives rises to cavity enhanced anti-Stokes scattering of  thermomechanical sideband around the cavity resonance.
In the weak coupling regime, we can linearize the mechanical displacement $\hat{b}\rightarrow\bar{b}+\delta \hat{b}$.
 In a resolved-sideband system, i.e. $\kappa \ll \Omega_m$, we can obtain the quantum Langevin equations for the field fluctuations   in the rotating frame of the cooling tone.
Within the rotating-wave approximation, we obtain,
\begin{equation}
\begin{aligned}
	\chi^{-1}\s{c,eff}(\Omega)\delta {\hat a}  = &
ig_0 \bar{a} \delta \hat{b}
+\sqrt{\kappa_a }\delta\hat a_d
+\sqrt{\kappa_{\s{ex}}}\delta\hat a\s{ex,in}
+\sqrt{\kappa_s}\delta\hat a\s{s,in}
\\
	\chi_m^{-1}(\Omega) \delta \hat{b}  =& ig_0 \bar{a}  \delta \hat{a}+\sqrt{\Gamma_m}\delta \hat{b}\s{in}
\end{aligned}
\label{eq:qleom}
\end{equation}
with mechanical susceptibility $\chi_m(\Omega)= 1/(\Gamma_m-i(\Omega-\Omega_m))$ and $\chi\s{c,eff}$ as defined in Eq.~\ref{eq:chiceff}.

Both coherent and incoherent optomechanical spectroscopy of the out-of-loop optical field can be adopted for the measurement of the mechanical motion~\cite{qiu_floquet_2019,qiu_laser_2020}.
In a coherent spectroscopy, a weak probing tone is generated from the strong pumping tone, e.g. via an electro-optical-modulator, with frequency separation set by the microwave tone.
The reflected light from the cavity is sent to the photodetector directly.
The photocurrent is demodulated at the microwave frequency, from which we can obtain a coherent cavity response with the probing tone sweeping across the optical micro-cavity resonance.
The cavity susceptibility is modified by the mechanical motion,
\begin{equation}
	\chi\s{c, om }(\Omega) = \frac{1}{\kappa\s{eff}/2 -i (\bar{\Delta}\s{eff} +\Omega )
		+\bar{n}_c g_0^2 \chi_m(\Omega)}.
	\label{eq:chicom}
\end{equation}
The mechanical motion leads to a destructive interference in the out-of-loop optical field.
The coherent optomechanical spectroscopy of the cavity field results in the optomechanical induced transparency~\cite{safavi-naeini_electromagnetically_2011,weis_optomechanically_2010}.
The dynamical back-action results in a modified mechanical susceptibility,
$\chi\s{m,eff}(\Omega) = 1/(\Gamma\s{eff}/2-i(\Omega-\Omega_m ))
$, with effective mechanical linewidth $\Gamma\s{eff}(\Omega)=\Gamma_m+\Gamma\s{opt}(\Omega)$ and optomechanical damping rate $\Gamma\s{opt}(\Omega)= \kappa\s{eff} \bar{n}_c g^2_0 |\chi\s{c,eff}(\Omega)|^2$.

For $g\s{th}>0$, i.e. temperature increase results in blue-shift of optical cavity frequency, the in-loop photon fluctuation is enhanced, i.e. $S\s{\delta I_a}(\Omega_m)>1$. This accordingly results in a decreased $\kappa\s{eff}$ and an increased optomechanical damping rate.
For $\Omega\s{m}\gg\gamma\s{th}$, such as in an optomechanical crystal cavity (OMC), the effective detuning change due to the dissipative feedback is negligible, i.e. $\bar{\Delta}\s{eff}\simeq\bar{\Delta}$. 
We note that, the photothermal effects can also change the mechanical susceptibility by heating the mechanical oscillators~\cite{deliberato_quantum_2011,metzger_cavity_2004,restrepo_classical_2011},
which is not considered here in our analysis.
The coherent dissipative dynamics has been shown to introduce Floquet dynamics in quantum measurement of mechanical motion~\cite{qiu_floquet_2019} and photothermal induced transparency~\cite{ma_photothermally_2020}.
Similar mechanism has been reported in a membrane in the middle system with active feedback~\cite{rossi_enhancing_2017a,rossi_normalmode_2018,zippilli_cavity_2018}.
This can give rises to enhanced sideband cooling and even normal-mode splitting in a weakly coupled optomechanical system. 

 The coupling to the dissipative reservoir can introduce excess perturbation to the mechanical motion.
In the resolved-sideband limit, the minimun final occupancy is obtained with $\bar{\Delta}\s{eff}=-\Omega_m$, and takes the form,
\begin{equation}
\bar{n}_f = \frac{\bar{n}\s{th} \Gamma_m+ 4 \bar{n}\s{l} \bar{n}_c g_0^2/\kappa\s{eff} }
{\Gamma_m+4\bar{n}_c g_0^2/\kappa\s{eff}},
\label{eq:nf}
\end{equation}
with 
\begin{equation}
\bar{n}_{l} = \kappa_a  \bar{n}^2_c \sigma^2\s{0}(\Omega_m)/\kappa\s{eff},
\end{equation}
the limit of sideband cooling in presence of the dissipative feedback.
The dissipative feedback gives rises to an effective backaction heating of the mechanical motion.
In the high cooperativity limit, i.e. $\Gamma\s{opt}\gg \bar{n}\s{th}\Gamma_m$, the final occupancy is limited by $\bar{n}_l$, due to the absorption induced dissipative feedback.

\begin{figure*}[tbh]
\centering
    \includegraphics[scale=1]{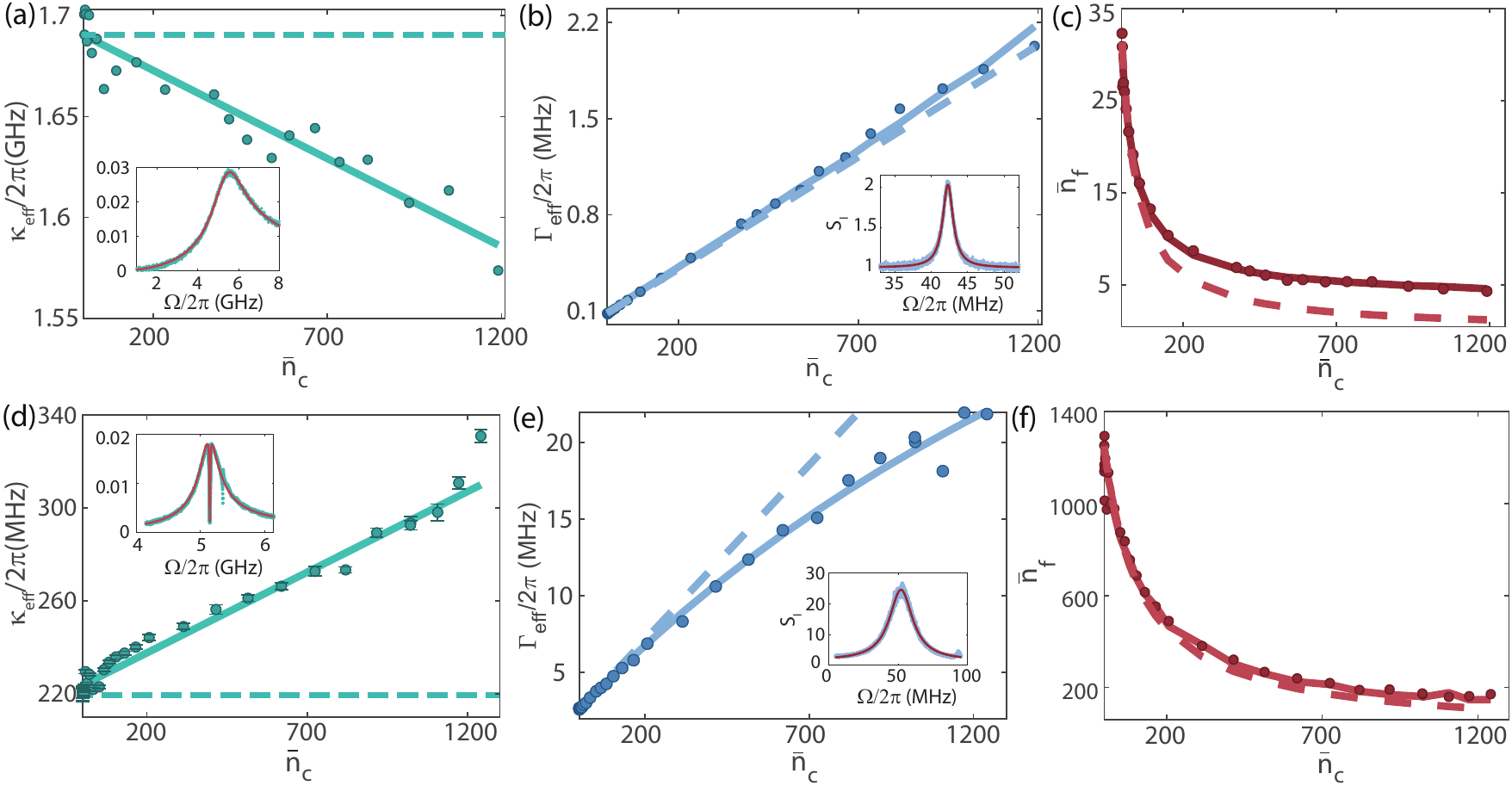}
  \caption{
  \textbf{Observation of dissipative feedback dynamics on cavity linewidth and optomechanical damping in optomechanical spectroscopy of two sideband-cooled OMCs under different conditions.}
   In (a-c), we show the results for an OMC of $\kappa/2\pi$ $\sim$1.7~GHz at $\sim$8K with a pressure $\sim$0.5mbar.
  (a) Fitted effective optical linewidth $\kappa\s{eff}$ (green  circles) vs. intracavity photon number $\bar{n}_c$ in the coherent cavity spectroscopy. Green curve shows a fitting curve incorporating the coherent dissipative dynamics, revealing a cavity linewidth that reduces with $\bar{n}_c$. The green dashed line shows the original cavity linewidth. The inset shows a typical fitting of coherent cavity response. 
  (b) Fitted effective mechanical linewidth $\Gamma\s{eff}$ (blue  circles) in the incoherent balanced heterodyne noise spectrum vs.  $\bar{n}_c$.
  Blue curve shows a fitting curve incorporating $\kappa\s{eff}$ due to coherent dissipative dynamics.
  Blue dashed curve shows a theoretical plot without the dissipative absorption.
  The inset shows a typical fitting of incoherent noise spectrum.
  (c) Mean phonon occupancy $\bar{n}_f$ (red  circles) from the incoherent  noise spectrum via mechanical noise thermometry vs. $\bar{n}_c$.
  The calibration is done using mechanical noise thermometry anchored at the lowest photon number.
  Red curve shows a fitting curve incorporating the effective optical linewidth change and linear absorption heating.
  Red dashed curve shows a theoretical plot without dissipative absorption.
  In (d-f), we show the results from an OMC of $\kappa/2\pi$ $\sim$220~MHz at room temperature and pressure, corresponding to the data in (a-c).
  Here an increase of $\kappa\s{eff}$ versus $\bar{n}_c$ is observed due to the dissipative feedback.
  In (e), the green dashed line shows the ideal theoretical effective mechanical linewidth without dissipative dynamics. 
  In (d), the error bars are from the fitting of the coherent cavity response, while in (a,b,c,e,f)  error bars are within the circles and thus not shown.
}
\label{fig:LTRT}
\end{figure*}

In an incoherent optomechanical spectroscopy,
the output field $\delta \hat{a} \s{ex,out}=\delta \hat{a}\s{ex,in}-\sqrt{\kappa\s{ex}}\delta \hat{a}$,
can be linearly measured by a quantum-limited balanced homodyne (heterodyne) detection by beating with a strong local oscillator.
In the balanced heterodyne detection (BHD), the local oscillator of frequency $\omega\s{LO}$ is placed on the blue side of the pump with a frequency separation close to the mechanical frequency.
In an ideal detection, the symmetrized power spectral density (PSD) of the output photocurrent from BHD
when normalized to the shot noise floor,
takes the form,
\begin{equation}
\begin{aligned}
S_I(\Omega+\Delta\s{LO}) =&
\frac{ \kappa\s{ex}  }{\kappa \s{eff}}\Gamma\s{opt}(\Omega+\Omega_m) \Gamma\s{eff}
 |\chi\s{m,eff}(\Omega+\Omega_m)|^2 \\
 &
\times  (\bar{n}_f-2\bar{n}_{l})
+ 1 + 4 \bar{n}_{l}\frac{\kappa\s{ex}}{\kappa\s{eff}},
\end{aligned}
\label{eq:SIhetero}
\end{equation}
assuming $\Delta\s{LO}=\Omega_m+\omega_L-\omega\s{LO}>0$.
Unanticipated \textit{excess classical correlation} between measurement imprecision and backaction arises due to the dissipative dynamics, which can even lead to \textit{noise squashing}, i.e. when $\bar{n}_f< 2 \bar{n}_l$.
In addition, the dissipative feedback results in an\emph{ increased noise floor,} i.e. $4\bar{n}_l \kappa\s{ex}/\kappa\s{eff}$, and an effective detection efficiency of $\kappa\s{ex}/\kappa\s{eff}$ as opposed to $\kappa\s{ex}/\kappa$.

We note that, both Eq.~\ref{eq:nf} and Eq.~\ref{eq:SIhetero} exhibit some similarities to an optomechanical system sideband-cooled by a laser with excess noise, which also introduces excess backaction heating and classical correlations~\cite{kippenberg_phase_2013,rabl_phasenoise_2009,safavi-naeini_laser_2013,sudhir_appearance_2017,weinstein_observation_2014}.
The incoherent dissipative dynamics is mediated via absorption induced cavity frequency fluctuation, by replacing the vacuum noise with excess backaction heating $\Gamma\s{opt}\bar{n}_l/\Gamma\s{eff}$ approximately quadratic to $\bar{n}_c$ (cf. Eq.~\ref{eq:nf}).
Surprisingly, even in absence of excess laser noise, the phonon occupancy can be \textit{underestimated} due to the excess classical correlation from the dissipative reservoir in the mechanical noise thermometry.
\section{Experimental Results}
We observe the dissipative dynamics on cavity linewidth, cooling rate and noise floor caused by photon absorption in an optomechanical spectroscopy of sideband-cooled optomechanical crystals (OMCs) with different optical losses at cryogenic temperature in vacuum and under ambient conditions respectively.
In a typical OMC, an optical mode $\sim$1550nm is coupled to a colocalized mechanical mode of $\Omega_m/2\pi$ $\sim$5~GHz at a vacuum optomechanical coupling rate of $g_0/2\pi$ $\sim$1~MHz.
Light is evanescently coupled into the OMC from a tapered optical fiber via a coupling waveguide, with a waveguide coupling efficiency $\eta\s{wg}$ ranging from $40\%$ to $60\%$.
The reflected light from the OMC is collected for optomechanical spectroscopy.
More details of the OMCs and the experimental setup are in the Appendix~\ref{app:exp}.

In the first set of measurements, we put an OMC device of $\kappa/2\pi$ $\sim$1.7~GHz (used in Ref.~\cite{qiu_floquet_2019,shomroni_optical_2019}) in a $^3$He buffer gas cryostat (Oxford Instruments HelioxTL). The experiments are performed at $\sim$8 K under vacuum (pressure $\sim$ 0.5 mbar), in contrast to previous experiments benifiting from the better thermalization due to gaseous $^3$He~\cite{qiu_floquet_2019,shomroni_optical_2019}.
In the coherent spectroscopy, the effective laser detuning $\bar{\Delta}\s{eff}$ is manually adjusted to $\sim -\Omega_m$, to achieve optimal sideband cooling. In Fig.~\ref{fig:LTRT}(a), the inferred $\kappa\s{eff}$ from the coherent spectroscopy vs. $\bar{n}_c$ are shown in green circles.
Typically, the nonlinear dynamics in the cavity, such as two-photon absorption (TPA) and free-carrier absorption (FCA), may increase the cavity linewidth due to the cavity enhancement~\cite{barclay_nonlinear_2005a}.
Counterintuitively, the inferred  $\kappa\s{eff}$ decreases as the cooling tone $\bar{n}_c$ increases, as shown in Fig.~\ref{fig:LTRT}(a).
We attribute this to the dissipative dynamics, where the optical mode is coupled to an effective dissipative reservoir, with $g\s{th}>0$ (blue-shift of cavity frequency due to temperature increase) for silicon OMC at 8K.
The optical absorption and decreased thermal conductivity of silicon at low temperature result in the optical heating of the OMC.
In Fig.~\ref{fig:LTRT}(a), the green curve shows
a linear fit using Eq.~\ref{eq:keff}, which results in a fitted $\kappa_a\sigma_0(\Omega_m)\simeq 2\pi\times 44 \mathrm{kHz}$.
At the highest power ($\bar{n}_c =1190$), \emph{the inferred $\kappa\s{eff}$ is decreased by $\sim7.5\%$ compared to the cavity linewidth $\kappa$}.
In Fig.~\ref{fig:LTRT}(b), the fitted  $\Gamma\s{eff}$ from the incoherent noise spectrum in the BHD are shown in blue  circles, with a light blue fitting curve incorporating $\kappa\s{eff}$, which results in $g_0/2\pi = 829 \mathrm{kHz}$ and $\Gamma_m/2\pi = 81\mathrm{kHz}$.
The effective damping rate is shown to be higher than the theoretical model without the absorption (the dashed blue curve), due to the increased optical susceptibility.
In Fig.\ref{fig:LTRT}(c), we show the calibrated final phonon occupancy $\bar{n}_f$ from the BHD using standard mechanical noise thermometry by anchoring the noise spectrum at the lowest power, i.e. $\bar{n}_c=1.4$.
A fitting curve with a theoretical model incorporating linear absorption heating is shown in the light red curve~\cite{qiu_floquet_2019}. The minimum phonon occupancy achieved is $\bar{n}_f \sim 4.3\pm0.1$ at $\bar{n}_c\sim 1200$.

In the second set of measurements, we perform an optomechanical spectroscopy of a sideband-cooled OMC of $\kappa/2\pi= 220\mathrm{MHz}$ (used in Ref.~\cite{qiu_laser_2020,shomroni_twotone_2019}) at room temperature and pressure.
In Fig.~\ref{fig:LTRT}(d), the fitted $\kappa\s{eff}$ vs. $\bar{n}_c$ are shown in green  circles from the coherent cavity response measurement, with $\bar{\Delta}\s{eff}\simeq-\Omega_m$.
At the highest powers, e.g. $\bar{n}_c\simeq1110$, \emph{the inferred $\kappa\s{eff}$ is increased by $\sim$ half of the original $\kappa$}.
Despite the low cavity loss, the dissipative dynamics arises due to the large thermorefractive coefficient and the cavity field enhancement.
In Fig.~\ref{fig:LTRT}(d), the light green curve shows a linear fitting using Eq.~\ref{eq:keff}, which results in a fitted $\kappa_a\sigma_0(\Omega_m) \simeq -2\pi \times 35\mathrm{kHz}$.
In Appendix~\ref{app:excess_noise}, we discuss additional possible nonlinear dynamics from TPA and FCA.
In Fig.1(e), we show the inferred $\Gamma\s{eff}$ from the BHD (blue  circles), which deviates from the linear scaling of $\bar{n}_c$ (blue dashed curve).
The light blue fitting curve incorporating $\kappa\s{eff}$ results in $g_0/2\pi = 1.12 \mathrm{MHz}$ and $\Gamma_m/2\pi = 2.56\mathrm{MHz}$.
In Fig.1(f), we show the calibrated $\bar{n}_f$ via mechanical noise thermometry, with the optomechanical damping rate incorporating $\kappa\s{eff}$.
The increased $\kappa\s{eff}$ at high powers results in an increased $\bar{n}_f$, with minimum $\bar{n}_f\sim 159\pm4$ at $\bar{n}_c=1107$.
\begin{figure}
\centering
    \includegraphics[scale=1]{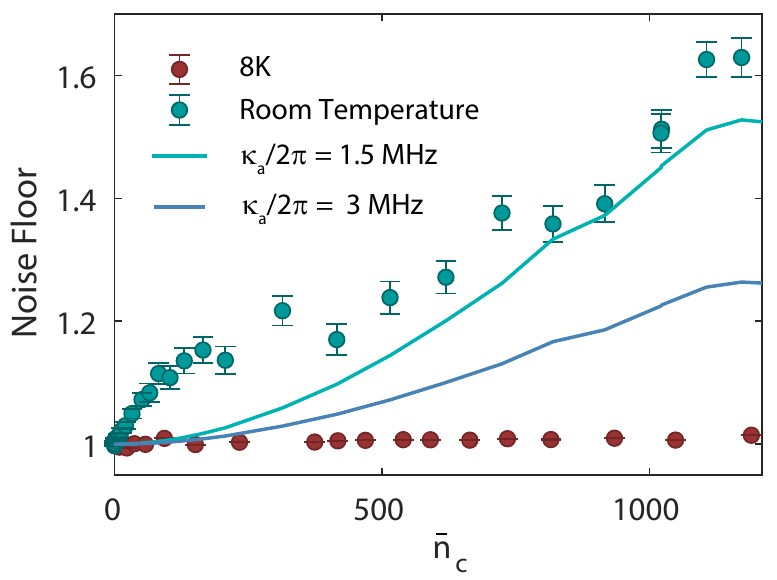}
  \caption{
  \textbf{Noise floor from the incoherent optomechanical spectroscopy of sideband-cooled OMCs at different intracavity photon numbers}. The noise floor is normalized to the shot noise floor.
   The red circles correspond to the OMC with
   $\kappa/2\pi\sim 1.7\mathrm{GHz}$ at $\sim$8K,
   while the green circles correspond to the OMC with $\kappa/2\pi\sim220\mathrm{MHz}$ under ambient conditions. 
   The green and blue curves correspond to the theoretical plots for $\kappa_a$ of $2\pi\times\mathrm{1.5MHz}$ and $2\pi\times\mathrm{3MHz}$, incorporating the inferred $\kappa\s{eff}$ and $\kappa_a \sigma_0(\Omega_m)$ from coherent cavity response.
}
\label{fig:BG}
\end{figure}

Figure~\ref{fig:BG} shows the noise floor in the incoherent noise spectrum from the two sets of measurements, normalized to the shot noise floor.
The red  circles correspond to the 8K measurements, where the noise floor changes negligibly with $\bar{n}_c$.
The green  circles correspond to the room temperature measurements, where the noise floor increases versus the optical power.
The large error bars are due to the large signal-to-noise ratio (SNR) of the thermomechanical sideband at room temperature.
We attribute the large noise floor increase to the dissipative dynamics, instead of the excess noise in the optical or electronic components in the setup, as detailed in  Appendix~\ref{app:excess_noise}.
Incorporating the fitted $\kappa\s{eff}$ and $\kappa_a \sigma_0(\Omega_m)$ from coherent response measurements at room temperature, we plot the theoretical curves of the noise floor assuming absorption rate $\kappa_a$ of $2\pi\times 3 \mathrm{MHz}$ (blue curve) and $2\pi\times 1.5 \mathrm{MHz}$ (green curve) in Fig.~\ref{fig:BG}.
In principle, $\kappa_a$ can be obtained from $\kappa\s{eff}$ and the noise floor from the coherent and incoherent spectroscopy, and the excess classical correlation in the mechanical noise thermometry can be estimated accordingly.
We observe large deviation of the increased noise floor at low powers, most likely due to the cavity frequency noise due to free-carrier absorption~\cite{barclay_nonlinear_2005a,hamerly_quantum_2015} as discussed in Appendix~\ref{app:excess_noise}.
Due to the complicated cavity dynamics, the excess noise correlations are not quantified in our measurements.

\section{Optical Kerr Squeezing}
The dissipative feedback arises in optomechanical spectroscopy due to photon absorption. 
Such photothermal effects manifest as a Kerr-type nonlinearity~\cite{qiu_floquet_2019}, where the cavity frequency shift is proportional to the intracavity photon number.
Kerr medium has been suggested as a quantum non-demolition (QND) device in a feedback loop to generate optical squeezing~\cite{shapiro_theory_1987,haus_theory_1986a}.
In such QND feedback, the intensity of in-loop optical field is measurement without quantum backaction, although phase fluctuations are added.
The dissipative feedback is fundementally different from a QND feedback, e.g. Kerr squeezing~\cite{shapiro_theory_1987,haus_theory_1986a}, as excess noise always arises in the out-of-loop field due to uncorrelated noise from the in-loop field detection, i.e. photon absorption.

Here we consider a micro-cavity with coexisting dissipative feedback via photothermal effects and Kerr nonlinearity. We can obtain the quantum Langevin equations of the field fluctuation in the frequency domain,
\begin{equation}
\begin{aligned}
	\chi^{-1}\s{c,0}(\Omega)\delta {\hat a}  =&
\left(\bar{n}_c \sigma_0(\Omega) \kappa_a -i \bar{n}_c g\s{Kerr}\right)
(\delta \hat{a}+\delta \hat{a}\dagg)\\
&+\sqrt{\kappa_a}\delta\hat a_d
 +\sqrt{\kappa_{\s{ex}}}\delta\hat a\s{ex,in}
+\sqrt{\kappa_s}\delta\hat a\s{s,in}
\end{aligned}
\end{equation}
where $g\s{Kerr}$ is the single photon nonlinear Kerr coupling rate~\cite{matsko_optical_2005} and $\chi\s{c,0}(\Omega)$ is given by Eq.~\ref{eq:chic0} with $\bar{\Delta}=\Delta -\left( g_a \kappa_a g\s{th}/\gamma\s{th} + g\s{Kerr}\right) \bar{n}_c$.
We can obtain the symmetrized PSD of the out-of-loop field quadrature, $\hat{X}_\theta  = \delta \hat{a}\s{ex,out}  e^{-i \theta}+ \delta \hat{a}\s{ex,out}\dagg  e^{i \theta}
$,
in a balanced homodyne detection, with LO of the pumping tone frequency and $\theta$ as the phase difference between the LO and the pump.
We focus on the simple case where the laser detuning $\bar{\Delta}=0$.
For Kerr squeezing, i.e. $g\s{th}=0$, the PSD takes the form,
\begin{widetext}
\begin{equation}
S\s{I,Kerr}(\Omega) = 1-
\frac{
16 \bar{n}_c\eta_c \kappa \sin(\theta)g\s{Kerr} \left(
 (\kappa^2+4 \Omega^2) \cos(\theta)-
4 \bar{n}_c \kappa \sin(\theta) g\s{Kerr}
\right)
}
{(\kappa^2+4\Omega^2)^2}.
\end{equation}
\end{widetext}
where we assume an ideal detection and $\eta_c=\kappa\s{ex}/\kappa$.
Such QND feedback can result in Kerr squeezing in the output field for an optical quadrature with minimum variance below the vacuum noise at an optimal angle, as detailed in Appendix~\ref{app:theory}.
In contrast, for dissipative dynamics only, i.e. $g\s{Kerr} = 0$, the PSD takes the form,
$S\s{I,a}(\Omega) = 1 + S^{\mathrm{ex}}\s{I,a}(\Omega)
$,
where the incoherent optical absorption always results in an excess noise, i.e. the photothermal noise,
\begin{equation}
S^{\mathrm{ex}}\s{I,a}(\Omega) =  \frac{16 \bar{n}_c  \kappa_a \kappa\s{ex} \sin(\theta)^2 g_a ^2 g\s{th}^2}{(\kappa^2+4 \Omega^2)(\Omega^2+\gamma\s{th}^2)}\geq0.
\label{eq:SIPTex}
\end{equation}
We note that, this coincides with earlier experimental and theoretical works in optical systems with intensity feedback,
where in-loop photocurrent can be squashed while the out-of-loop photocurrent always becomes anti-squashed~\cite{taubman_intensity_1995a,wiseman_inloop_1998,wiseman_squashed_1999}.
Kerr nonlinearity on the other hand, demonstrates a QND detection of the in-loop optical field, which enables the Kerr squeezing in the out-of-loop field, as expected~\cite{shapiro_theory_1987,haus_theory_1986a}.

When Kerr nonlinearity and dissipative dynamics coexist, the total PSD is given by $S_I(\Omega) = S\s{I,Kerr}(\Omega)+S^{\mathrm{ex}}\s{I,tot}(\Omega)$, with total excess noise $S^{\mathrm{ex}}\s{I,tot} (\Omega)= S^{\mathrm{ex}}\s{I,a}(\Omega)+S^{\mathrm{ex}}\s{I,c}(\Omega)$ due to the dissipative dynamics.
The coherent dissipative dynamics gives arises to,
\begin{equation}
S^{\mathrm{ex}}\s{I,c}(\Omega)=\frac{64 \bar{n}^2_c \kappa_a \kappa\s{ex}  g_a g\s{Kerr} g_{\text{th}} \sin ^2(\theta ) \left(\kappa  \gamma _{\text{th}}-2 \Omega ^2\right)}{\left(\kappa ^2+4 \Omega ^2\right)^2 \left(\gamma _{\text{th}}^2+\Omega ^2\right)},
\end{equation}
We note that, when
$g_a g_{\text{th}} \left(\kappa ^2+4 \Omega ^2\right)<g\s{Kerr} \left(8 \Omega ^2-4 \kappa  \gamma _{\text{th}}\right)$,
the total excess noise $ S^{\mathrm{ex}}\s{I,tot}<0$, which \emph{opens an interesting regime for dissipation improved Kerr squeezing}.
Such regime is  independent of the intracavity photon number and squeezing angle.

\begin{figure}
\centering
    \includegraphics[scale=1]{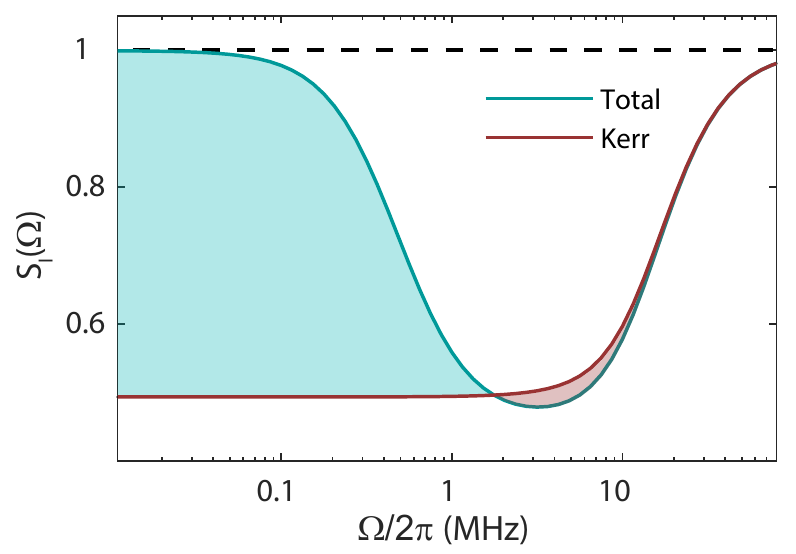}
  \caption{\textbf{Estimated Kerr squeezing with dissipative dynamics at optimal angles in a $\textrm{Si\s{3}N\s{4}}$  micro-cavity.}
  The noise spectral density is normalized to the shot noise floor (black dashed line).
  The red curve corresponds to the case where the dissipative dynamics is absent, while the green curve corresponds to the case where Kerr nonlinearity and dissipative dynamics coexist. The green and red shade areas correspond to the excess noise and improved squeezing due to dissipative feedback.
   (Parameters: $\kappa/2\pi = 15\mathrm{MHz}$, $\kappa\s{ex}/2\pi = 8\mathrm{MHz}$, $\kappa_a/2\pi = 6 \mathrm{MHz}$, $\kappa_s/2\pi  = 1 \mathrm{MHz}$, $\gamma\s{th}/2\pi = 20 \mathrm{kHz}$, $g_a g\s{th}/2\pi = - 0.05 \mathrm{Hz}$, $g\s{Kerr}/2\pi = -0.5 \mathrm{Hz}$, and $\bar{n}_c=10^7$)
  }
\label{fig:kerrsq}
\end{figure}

In Fig.~\ref{fig:kerrsq}, we show the theoretical curves of estimated Kerr squeezing in a state-of-art $\textrm{Si\s{3}N\s{4}}$ micro-ring resonators at optimal squeezing angles  with realistic parameters~\cite{liu_highyield_2020}.
Kerr nonlinearity in such micro-ring resonator can lead to frequency comb and dissipative soliton generation at low pumping powers ($\sim$mW)~\cite{gaeta_photonicchipbased_2019}, due to the extremely low optical loss ($\kappa/2\pi<20\mathrm{MHz}$). The red curve includes only Kerr nonlinearity while the green curve considers both the Kerr nonlinearity and the dissipative dynamics.  As shown in Fig.~\ref{fig:kerrsq}, the green curve exhibits excess noise at low frequencies due to the incoherent dissipative dynamics, while shows slight improvement of Kerr squeezing at frequencies between 2MHz$\sim$10MHz due to the coherent dissipative dynamics.
In practice, the Kerr squeezing is limited by the detection efficiency and the thermorefractive noise~\cite{braginsky_thermorefractive_2000,huang_thermorefractive_2019,panuski_fundamental_2020}.

\section{Conclusion}
We observe the dissipative dynamics resulting from photon absorption in quantum optomechanical spectroscopy of sideband-cooled optomechanical crystal cavities under different conditions and develop a theoretical model for our experimental observations.
Such dissipative dynamics comes from a  dissipative quantum  feedback to the optical cavity.
The in-loop detection of the optical absorption field results in a modified optical susceptibility, and results in excess noise correlations in the quantum measurement of mechanical motion.
Such dissipative feedback differs from the quantum non-demolition feedback, e.g. Kerr squeezing.
The observed dynamics is applicable to any dissipative feedback in the cavities of different bandwidths, such as TPA, FCA, and photorefractive effects~\cite{li_photonlevel_2019}.
It offers crucial understanding to cavity spectroscopy in both optical and superconducting microwave micro-cavities, such as noise thermometry, quantum enhanced sensing, and microwave-optical frequency conversions~\cite{sahu_quantum-enabled_2021,holzgrafe_cavity_2020}. 
While our study has experimentally considered the optical domain, it is worth pointing out, that our results also apply to the microwave domain, as studied in circuit optomechanics or used in superconducting circuits based research~\cite{aspelmeyer_cavity_2014,blais_circuit_2021}. While it is well known that TLS can lead to a drive power dependent quality factor of superconducting microwave cavities, the described phenomena may be applicable to mechanisms  which result in power or temperature dependent cavity frequencies, e.g. due to quasiparticles~\cite{pop_coherent_2014,bespalov_theoretical_2016,gustavsson_suppressing_2016,tai_coherence_2021}.
The dissipative feedback can also be exploited for future dissipation engineering, such as enhanced sideband cooling and Kerr squeezing, and nonreciprocal photonic devices.
\begin{acknowledgments}
L.Q. acknowledges fruitful discussions with D. Vitali, R. Schnabel, P.K. Lam, A. Nunnenkamp, and D. Malz. This work is supported by the EUH2020 research and innovation programme under Grant No. 732894 (FET Proactive HOT), and the European Research Council through Grant No. 835329 (ExCOM-cCEO). This work was further supported by Swiss National Science Foundation under Grant Agreements No. 185870 (Ambizione) and No. 204927. Samples were fabricated at the Center of MicroNanoTechnology (CMi) at EPFL and the Binnig and Rohrer Nanotechnology Center at IBM Research-Zurich.
\end{acknowledgments}
\section*{Data Availability Statement}
All data and analysis files will be made available via \texttt{zenodo.org} upon publication.

\appendix
\section{Theory}\label{app:theory}
\subsection*{Optomechanical Spectroscopy}
For laser sideband cooling in a resolved-sideband optomechanical system coupled to a dissipative reservoir,
the quantum Langevin equation takes the form,
\begin{equation}
\begin{aligned}
	\chi^{-1}\s{c,eff}(\Omega)\delta {\hat a}  = &
ig_0 \bar{a} \delta \hat{b}
+\sqrt{\kappa_a }\delta\hat a_d
+\sqrt{\kappa_{\s{ex}}}\delta\hat a\s{ex,in}
+\sqrt{\kappa_s}\delta\hat a\s{s,in}
\\
	\chi_m^{-1}(\Omega) \delta \hat{b}  =& ig_0 \bar{a}  \delta \hat{a}+\sqrt{\Gamma_m}\delta \hat{b}\s{in}
\end{aligned}
\label{eq:qleomSI}
\end{equation}
and the noise operators satisfy the following correlations,
\begin{equation}
\begin{aligned}
\left<\delta \hat{a}\s{ex,in}\dagg(\Omega) \delta \hat{a}\s{ex,in}(\Omega^\prime)\right>=&  0 \\
\left<\delta \hat{a}\s{ex,in}(\Omega) \delta \hat{a}\dagg \s{ex,in}(\Omega^\prime)\right>
=&  \delta(\Omega+\Omega^\prime) 2\pi \\
\left<\delta \hat{a}\s{0,in}\dagg(\Omega) \delta \hat{a}\s{0,in}(\Omega^\prime)\right>=&  0 \\
\left<\delta \hat{a}\s{0,in}(\Omega) \delta \hat{a}\dagg \s{0,in}(\Omega^\prime)\right>=&  \delta(\Omega+\Omega^\prime) 2\pi \\
\left<\delta \hat{b}\s{in}\dagg(\Omega) \delta \hat{b}\s{in}(\Omega^\prime)\right>=&  \bar{n}\s{th}\delta(\Omega+\Omega^\prime) 2\pi \\
\left<\delta \hat{b}\s{in}(\Omega) \delta \hat{b}\dagg \s{in}(\Omega^\prime)\right>=& (\bar{n}\s{th}+1) \delta(\Omega+\Omega^\prime) 2\pi
\end{aligned}
\end{equation}
In addition, the noise operator for the dissipative reservoir satisfies the following correlations,
\begin{equation}
\begin{aligned}
\left<\delta \hat{a}_d\dagg(\Omega) \delta \hat{a}_d(\Omega^\prime)\right>=&
-\sigma_d(\Omega)  \sigma_d( \Omega^\prime)
 \delta(\Omega+\Omega^\prime) 2\pi\\
\left<\delta \hat{a}_d(\Omega) \delta \hat{a}\dagg_d(\Omega^\prime)\right>=& (1-\sigma_d(\Omega))(1+\sigma_d(\Omega^\prime))
\delta(\Omega+\Omega^\prime) 2\pi\\
\left<\delta \hat{a}_d(\Omega) \delta \hat{a}_d(\Omega^\prime)\right>=&
-(1-\sigma_d(\Omega)) \sigma_d(\Omega^\prime)
 \delta(\Omega+\Omega^\prime) 2\pi\\
\left<\delta \hat{a}\dagg_d(\Omega) \delta \hat{a}\dagg_d(\Omega^\prime)\right>=& \sigma_d(\Omega)(1+\sigma_d(\Omega^\prime)) \delta(\Omega+\Omega^\prime) 2\pi.
\end{aligned}
\end{equation}
We can obtain the noise spectrum of the mechanical oscillator,
\begin{equation}
\begin{aligned}
S\s{\hat{b}\hat{b}}(\Omega) &=\int_{-\infty}^{\infty}\left<\delta \hat{b}\dagg(t) \delta \hat{b}(t^\prime)\right> e^{i \Omega t} dt\\
& =\left(\bar{n}\s{th} \Gamma_m+ \Gamma\s{+}\right) |\chi\s{eff}( -\Omega)|^2
\end{aligned}
\end{equation}
where $\Gamma\s{+}= \bar{n}_c g^2_0    |\chi_c(\Omega_m)|^2\kappa_a \sigma^2_d$, and
\begin{equation}
\begin{aligned}
S\s{\hat{b}\dagg\hat{b}\dagg}(\Omega) &=\int_{-\infty}^{\infty}\left<\delta \hat{b}(t) \delta \hat{b}\dagg(t^\prime)\right> e^{i \Omega t} dt\\
& =\left((\bar{n}\s{th}+1) \Gamma_m + \Gamma\s{-} \right) |\chi\s{eff}(\Omega)|^2
\end{aligned}
\end{equation}
where
$\Gamma\s{-} =  \bar{n}_c g^2_0 |\chi_c(\Omega_m)|^2 (\kappa\s{eff}+\kappa_a \sigma_d ^2)$.
The two-sided mechanical noise spectrum can be obtained,
\begin{equation}
\begin{aligned}
S\s{\hat{x} \hat{x} }(\Omega)/x^2\s{zpf} &=S\s{\hat{b}\dagg\hat{b}\dagg}(\Omega)+S\s{\hat{b}\hat{b}}(\Omega)\\
& =\Gamma\s{eff} \left[|\chi\s{eff}(\Omega)|^2 (\bar{n}_f+1)+  |\chi\s{eff}(-\Omega)|^2 \bar{n}_f\right],
\end{aligned}
\end{equation}
where the final occupancy takes the form,
\begin{equation}
\bar{n}_f = \frac{\bar{n}\s{th} \Gamma_m+ \Gamma\s{opt} \bar{n}_{l}}{\Gamma\s{eff}},
\label{eq:nfSI}
\end{equation}
with $\Gamma\s{opt}= \kappa\s{eff} \bar{n}_c g^2_0 |\chi\s{c,eff}(\Omega)|^2$ and $\bar{n}_{l} = \kappa_a  \bar{n}^2_c \sigma^2\s{0}/\kappa\s{eff}$.
The coupling to the dissipative reservoir results in excess backaction heating of the mechanical oscillator.
In a balanced heterodyne measurement, the output field is beating with a strong LO of frequency $\omega\s{LO}$,
which is placed on the blue side of the pumping tone with a frequency separation close to the mechanical frequency.
The symmetrized power spectral density (PSD) of the photocurrent from the BHD is given by,
$S_I(\Omega) =\frac{1}{2} \int_{-\infty}^{\infty}\langle\overline{\left\{\delta I\left(t+t^{\prime}\right), \delta I\left(t^{\prime}\right)\right\}}\rangle e^{i \Omega t} d t$.
In practice, there are other losses in the photodetection, such as in the fiber or the photodetection,
resulting in a limited external detection efficiency $\eta\s{ex}$.
Under optimal sideband cooling condition, i.e. $\bar{\Delta}\s{eff}=-\Omega_m$,
the photocurrent PSD, when normalized to the shot noise floor, is given by,
\begin{equation}
\begin{aligned}
S_I(\Omega+\Delta\s{LO}) =&
\eta\s{ex} \frac{4 \kappa\s{ex} \bar{n}_c g^2_0}{\kappa^2\s{eff}}\Gamma\s{eff}  |\chi\s{m,eff}(\Omega+\Omega_m)|^2
\\
&
 \times (\bar{n}_f-2\bar{n}_{l})
  + 1 + 4 \eta\s{ex} \bar{n}_{l}\frac{\kappa\s{ex}}{\kappa\s{eff}}
  \end{aligned},
\label{eq:SIpt}
\end{equation}
with $\Delta\s{LO}=\Omega_m+\omega_L-\omega\s{LO}>0$ and $\Gamma\s{eff} = \Gamma_m + 4 \bar{n}_c g^2_0/\kappa\s{eff}$.
At low cooling power, the signal-to-noise ratio of the thermomechanical sideband is given by,
\begin{equation}
\mathrm{SNR} = \frac{16 \eta\s{ex} \bar{n}_c g^2_0\kappa\s{ex}k_B T }{\kappa^2 \Gamma_m \hbar \Omega_m},
\end{equation}
with bath temperature T and negligible dynamical backaction.
In the fitting of $\bar{n}_c$ dependent $\Gamma\s{eff}$ in the main text, the inferred $\kappa\s{eff}$ is incorporated,  with $g_0$ and $\Gamma_m$ as free fitting parameters.
The mechanical noise thermometry can be performed by anchoring the phonon occupancy to the lowest power via Eq.~\ref{eq:SIpt}, incorporating the inferred $\kappa\s{eff}$ in the optomechanical damping rate.
The external detection efficiency $\eta\s{ex}$ can be obtained accordingly, and is estimated to be $\sim 0.15$ in the room temperature experiments.
The dissipative dynamics results in an increased noise floor of
\begin{equation}
BG\s{ex} = 4 \eta\s{ex} \bar{n}\s{l}\frac{\kappa\s{ex}}{\kappa\s{eff}}
=  \frac{4 \eta\s{ex} \kappa_a \kappa\s{ex} \bar{n}^2_c \sigma^2_0}{\kappa\s{eff}^2}.
\label{eq:BGex}
\end{equation}
However, Eq.~\ref{eq:BGex} is not sufficient to fully capture the observed noise increase, which may be partially from the cavity frequency noise due to free-carrier absorption~\cite{barclay_nonlinear_2005a,hamerly_quantum_2015}.
For the fitting of $\bar{n}_c$ dependent $\bar{n}_f$ in the main text, we incorporate a linear optical absorption for the mechanical motion, as in Ref.~\cite{qiu_floquet_2019}.

\subsection*{Optical Kerr Squeezing}
The Kerr nonlinearity coupling rate in the main text can be estimated through~\cite{matsko_optical_2005}
\begin{equation}
  g\s{Kerr} = -\omega_c\frac{n_2}{n_0}\frac{\hbar\omega_cc}{V\s{mode}n_0}<0,
\end{equation}
where $n_0$ is the linear refractive index, $n_2$ the Kerr coefficient, $V\s{mode}$ the mode volume of the cavity, and $c$ the speed of light.
The output field quadrature can be linearly measured in a balanced homodyne detection, by beating with a strong LO of the pumping tone frequency.
This results in the photocurrent fluctuation,
$\delta \hat{I} (t) = |\bar{a}\s{LO}| \hat{X}_\theta
$,
where $\theta$ is the phase difference between the LO and pumping tone.
The symmetrized noise power spectral density (PSD) of the output photocurrent, when normalized to the shot-noise floor, is
\begin{equation}
\begin{aligned}
S_I(\Omega) &=\left\langle \hat{X}_{\theta}(-\Omega) \hat{X}_{\theta}(\Omega)\right\rangle_{s} \\
&=\frac{1}{2}\left(\left\langle \hat{X}_{\theta}(-\Omega) \hat{X}_{\theta}(\Omega)\right\rangle+\left\langle \hat{X}_{\theta}(\Omega) \hat{X}_{\theta}(-\Omega)\right\rangle\right)
\end{aligned}
\end{equation}
The full expression of the PSD can be obtained analytically, which is too complicated to be presented here.
We focus on the simple case where the laser detuning $\bar{\Delta}=0$.
In absence of the dissipative dynamics, i.e. $g\s{th}=0$, the PSD takes the form,
\begin{widetext}
\begin{equation}
S\s{I,Kerr}(\Omega) = 1-
\frac{
16 \bar{n}_c \eta\s{ex} \eta_c \kappa \sin(\theta)g\s{Kerr} \left(
 (\kappa^2+4 \Omega^2) \cos(\theta)-
4 \bar{n}_c \kappa \sin(\theta) g\s{Kerr}
\right)
}
{(\kappa^2+4\Omega^2)^2}.
\end{equation}
\end{widetext}
with cavity coupling efficiency $\eta_c=\kappa\s{ex}/\kappa$ and external detection efficiency $\eta\s{ex}$.
The minimum PSD,
\begin{equation}
S\s{I,Kerr}^{\mathrm{min}}(\Omega)=1-\frac{2 \eta\s{ex} \eta_c }{\sqrt{\left( (\kappa ^2+4 \Omega ^2)/4 \bar{n}_c \kappa  g\s{Kerr}\right)^2+1}+1},
\end{equation}
which is limited by the cavity linewidth and total detection efficiency $\eta\s{ex} \eta_c$,
can be obtained at
$\theta\s{opt}(\Omega)= \tan ^{-1}\left(
(\kappa ^2+4 \Omega ^2)/4 \bar{n}_c \kappa  g\s{Kerr}
\right)/2
$.
In absence of the Kerr nonlinearity, i.e. $g\s{Kerr} = 0$, the PSD takes the form
$S\s{I,a}(\Omega) = 1 + S^{\mathrm{ex}}\s{I,a}(\Omega)
$,
where the incoherent dissipative dynamics always results in an excess noise,
\begin{equation}
S^{\mathrm{ex}}\s{I,a}(\Omega) =  \frac{16 \bar{n}_c \eta\s{ex} \eta_a  \eta_c \kappa^2 \sin(\theta)^2 g_a ^2 g\s{th}^2}{(\kappa^2+4 \Omega^2)(\Omega^2+\gamma\s{th}^2)}\geq0,
\end{equation}
with $\eta\s{a}=\kappa_a/\kappa$.

In most optical systems, the Kerr nonlinearity and dissipative dynamics coexist, the total PSD is given by
\begin{equation}
S_I(\Omega) = S\s{I,Kerr}(\Omega)+S^{\mathrm{ex}}\s{I,tot}(\Omega),
\label{eq:SISq}
\end{equation}
with total excess noise $S^{\mathrm{ex}}\s{I,tot} (\Omega)= S^{\mathrm{ex}}\s{I,a}(\Omega)+S^{\mathrm{ex}}\s{I,c}(\Omega)$ due to the absorption.
The \emph{coherent dissipative dynamics} gives rise to,
\begin{equation}
S^{\mathrm{ex}}\s{I,c}(\Omega)=\frac{64 \bar{n}^2_c \eta\s{ex} \eta_a \eta_c  \kappa ^2 g_a g\s{Kerr} g_{\text{th}} \sin ^2(\theta ) \left(\kappa  \gamma _{\text{th}}-2 \Omega ^2\right)}{\left(\kappa ^2+4 \Omega ^2\right)^2 \left(\gamma _{\text{th}}^2+\Omega ^2\right)}.
\end{equation}
The minimum quadrature can be obtained accordingly at the frequency dependent optimal squeezing angle,
\begin{equation}
\theta\s{opt}(\Omega)
= \begin{cases}
      \frac{\arctan(-B/A)}{2} & A< 0 \\
      \frac{\arctan(-B/A)+\pi}{2} & A> 0
   \end{cases},
\end{equation}
where
\begin{widetext}
\begin{equation}
\begin{aligned}
A &=\bar{n}_c \kappa  \left(4 \eta_a g_a g\s{Kerr} g_{\text{th}} \left(2 \Omega ^2-\kappa  \gamma _{\text{th}}\right)-\eta_a g_a^2 g_{\text{th}}^2 \left(\kappa ^2+4 \Omega ^2\right)-4 g\s{Kerr}^2 \left(\gamma _{\text{th}}^2+\Omega ^2\right)\right)/g\s{Kerr}\\
B &= \left(\kappa ^2+4 \Omega ^2\right) \left(\gamma _{\text{th}}^2+\Omega ^2\right)>0.
\end{aligned}
\end{equation}
\end{widetext}

We note that, when
$g_a g_{\text{th}} \left(\kappa ^2+4 \Omega ^2\right)<g\s{Kerr} \left(8 \Omega ^2-4 \kappa  \gamma _{\text{th}}\right)$,
the total excess noise $ S^{\mathrm{ex}}\s{I,tot}<0$.
This brings about an interesting regime for dissipation assisted Kerr squeezing, which is  independent from the intracavity photon number and squeezing angle.

\section{Sample and Experimental Details}\label{app:exp}
Both OMCs in the main text are fabricated on a silicon-on-insulator wafer (Soitec) with a top-silicon device-layer thickness of $220\mathrm{nm}$ and a buried-oxide layer thickness of $2\mathrm{\mu m}$.
The OMC with $\kappa/2\pi = 1.6\mathrm{GHz}$,
is patterned by electron beam lithography (EBL)
using ZEP520A (100\%) as a positive resist.
The pattern is transferred to the device layer following a reactive ion etching using $\mathrm{SF_6}/\mathrm{C_4 F_8}$ plasma~\cite{qiu_floquet_2019,shomroni_optical_2019}.
The OMC of $\kappa/2\pi=220\mathrm{MHz}$,
is patterned by using EBL
using 4\% hydrogen silsesquioxane (HSQ) as a negative resist.
Pattern transfer into the device layer is accomplished by inductively-coupled-plasma reactive ion etching (ICP-RIE) using $\mathrm{HBr}/\mathrm{O}_2$~\cite{qiu_laser_2020,shomroni_twotone_2019}.
To permit input/output coupling with a tapered fiber, an additional photolithography step is performed for both OMCs followed by reactive ion etching (RIE) with a mixture of $\mathrm{SF_6}$ and $\mathrm{C_4 F_8}$ to create a mesa structure.
After resist removal, the buried oxide layer is partially removed in 10\% hydrofluoric acid
to create free-standing devices.
A Piranha (a mixture of sulfuric acid and hydrogen peroxide)
cleaning step is performed to remove organic residues.
In the end, the sample is dipped into 2\% hydrofluoric acid to terminate the silicon surface with hydrogen atoms.

The characterized parameters for the two OMCs under different measurement conditions are listed in Tab.~\ref{tab:params}.
\begin{table}
\caption{\textbf{Detailed characterized parameters of the two tested OMC device D1 and D2.} D1 is measured at $\sim$ 8K with a pressure of 0.5mbar, while D2 is measured at ambient conditions. }
\begin{tabular}{ |c|c|c|c|c|c|c|c| }
 \hline

 & $\lambda$ (nm)
 & $\kappa/2\pi$
 & $\kappa\s{ex}/2\pi$
 & $\eta\s{wg}$
 & $\Omega_m/2\pi$
 & $\Gamma_m/2\pi$
 & $g_0/2\pi$
 \\
 \hline
 D1
 & 1540
 & 1.7GHz
 & 0.5GHz
 & 0.52
 & 5.3GHz
 & 81kHz
 & 829kHz
 \\
 \hline
 D2
 & 1540
 & 220MHz
 & 73MHz
 & 0.5
 & 5.14GHz
 & 2.56MHz
 & 1.12MHz\\
 \hline
\end{tabular}
 \label{tab:params}
\end{table}
The measurement of the OMC of $\kappa/2\pi = 1.6 \mathrm{GHz}$ is performed in a $^3$He buffer gas cryostat (Oxford Instruments HelioxTL).
As shown in our previous experiments, the gaseous $^3$He improves the thermalization of the silicon OMC significantly.
In this experiment, the pressure of buffer gas is reduced to $\sim$0.5 mbar with the cryostat temperature stabilized around 8K, resulting in significant optical absorption heating.

\begin{figure}
\centering
    \includegraphics[scale=1.]{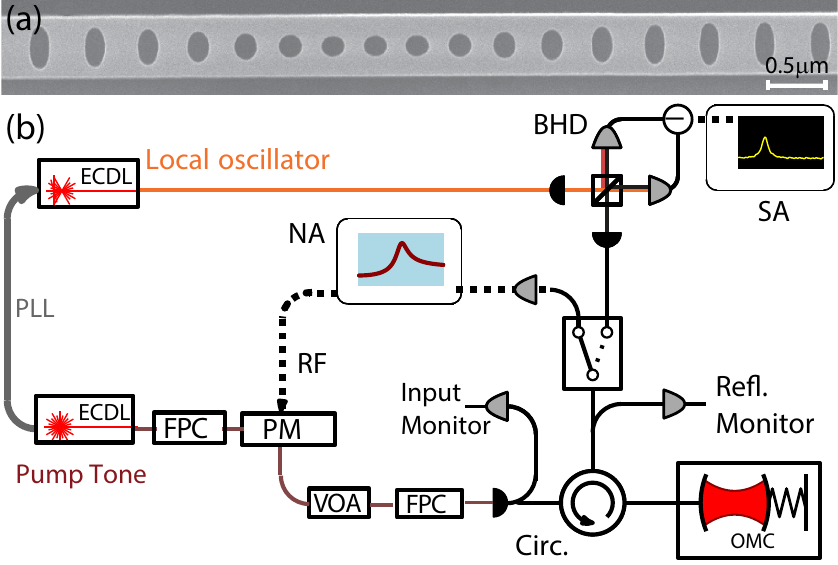}
  \caption{\textbf{Experimental setup for optomechanical spectroscopy.} 
  	(a) SEM image of the silicon optomechanical crystal cavity.
  	(b) Detailed experimental setup. ECDL, external-cavity diode lasers; FPC, fiber polarization controller; PM, phase modulator; VOA, variable optical attenuator; BHD, balanced heterodyne detector; SA, spectrum analyzer; NA, network analyzer; PLL, phase-locked loop.
  }
\label{fig:setup}
\end{figure}
In Fig.~\ref{fig:setup}, we show the experimental setup for the optomechanical spectroscopy.
Light from the ECDL passes through a phase modulator. A pair of weak probing tones is generated, with frequency separation set by a microwave tone from a vector network analyzer (VNA).
Light is coupled into the OMC with a tapered optical fiber.
The reflected light from the OMC is collected for optomechanical spectroscopy.
In the coherent spectroscopy, the photocurrent of the reflected light is sent to the VNA and demodulated at the microwave frequency, resulting in a coherent cavity response.
In the incoherent spectroscopy, the reflected light is sent to a balanced heterodyne detection setup by beating with a strong local oscillator, which is phase locked to the blue side of the cooling tone. The photocurrent from the BHD is sent to a spectrum analyzer, enabling a quantum-limited detection of the thermomechanical sideband.

\begin{figure}
\centering
    \includegraphics[scale=1]{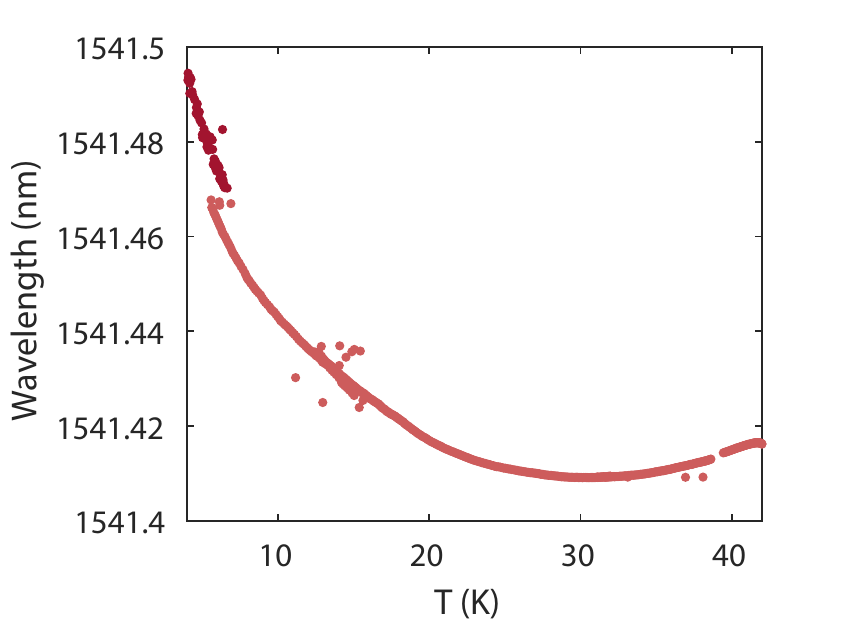}
  \caption{\textbf{Optical cavity resonant wavelength vs. cryostat temperature of a silicon OMC. }
  The measurement is performed with the OMC with $\kappa/2\pi = 1.6\mathrm{GHz}$ at pressure $\sim100\mathrm{mbar}$ with  $\bar{n}_c<5$ by monitoring the coherent cavity response. The slight color difference at temperature below 8K corresponds to two different sets of measurements.
  }
\label{fig:lambdaT}
\end{figure}
In Fig.~\ref{fig:lambdaT}, the optical cavity resonant wavelength vs. the cryostat temperature of a silicon OMC is shown. For temperature above 30K, $g\s{th}<0$, with a coefficient few orders of magnitude than room temperature. For temperature below 30K, $g\s{th}>0$, as opposed to room temperature. This enables stable strong pumping at the cavity red sideband in sideband cooling.

In practice, due to different fabrication process the absorption rate $\kappa_a$ may differ qualitatively, thus is very difficult to quantify, such as using cavity enhanced photothermal spectroscopy. 
In principle, $\kappa_a$ and $\sigma_0(\Omega_m)$ can be jointly fitted from the effective linewidth change (c.f. Eq.~\ref{eq:keff}) and the excess noise floor (c.f. Eq.~\ref{eq:BGex}), as they scale differently to the intracavity photon number. 
More specifically, we estimate $\kappa_a$ at room temperature of $\sim2\pi\times 1.5 \rm{MHz}$, corresponding to $\sigma_0(\Omega)$  of $\sim0.02$. 
We note that, the TPA and FCA in silicon OMC may also result in cavity linewidth increase and frequency blue shift, which is discussed in Appendix~\ref{app:excess_noise}. 
At cryogenic temperatuers~(8K), we estimate $\kappa_a  \sigma_0(\Omega_m)$ of $2\pi\times44\rm{kHz}$. Because of the relatively small noise floor change~($1\%$), we could not reliably quantify the absorption rate of such OMC.
We note that, the esimated $\sigma_0(\Omega)$ at room temperature and cryogenic temperature are of the same magnitude while of different sign. As the temperature dependent cavity frequency shift ($g_{\rm{th}}$) are different by almost two orders of magnitude between room temperature and cryogenic temperature, the absorption loss rate of the OMC at cryogenic temperature may be  $ \sim100\rm{MHz}$. Such different absorption rates of the two OMCs may explain our previous different optimal ground cooling results, i.e. with final phonon occupancy of 1.5 and 0.1 respectively~\cite{qiu_floquet_2019,qiu_laser_2020}.

\section{Thermal Response Simulations}\label{app:simulation}
In the theoretical treatment of the temperature dynamics in the main text, we neglect the geometric dependence.
In practice, the thermal response can be rather complex as dissipative absorption in the microcavities can dissipate via different mechanisms, such as radiation, conduction, and convection.
\begin{figure}
\centering
    \includegraphics[scale=1]{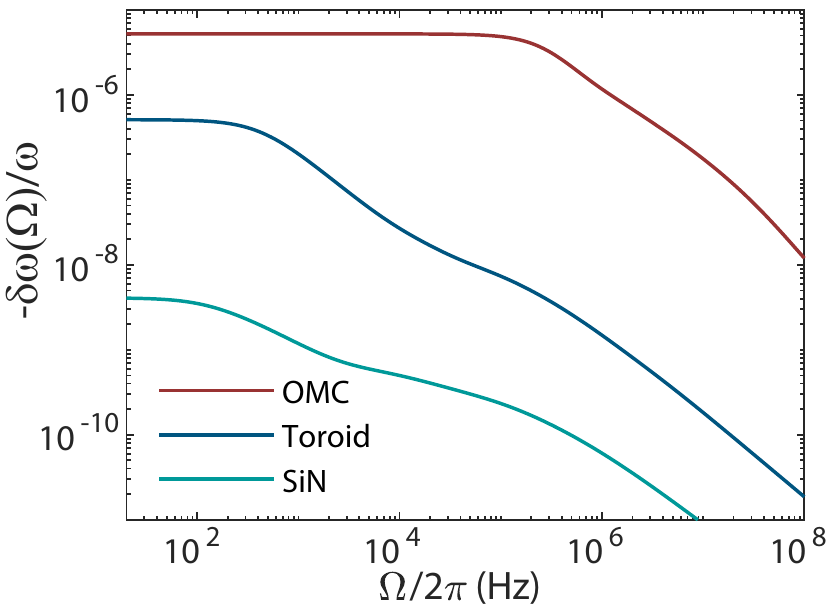}
  \caption{\textbf{Simulated cavity frequency response of different micro-cavities from finite element simulations at room temperature, including a silicon OMC~\cite{chan_optimized_2012}, silica micro-toroid~\cite{verhagen_quantumcoherent_2012}, and SiN ring resonator~\cite{liu_highyield_2020}. The detailed parameters in the FEM simulation are listed in Tab.~\ref{tab:coef}.}
  }
\label{fig:responses}
\end{figure}
The thermal response can be obtained  with finite element methods (FEM) simulations.
We first simulate the electric field distribution of the optical mode $\vec{E}(\vec{r})$.
The bulk absorption heating \begin{equation}
P(\vec{r})=P_\mathrm{abs}\epsilon_0\vec{E}(\vec{r})\times\hat{\epsilon}\vec{E}(\vec{r})/W_E,
\end{equation}
can be added as a heat source in the heat transfer model,
with $\hat{\epsilon}=\hat{\epsilon}_0+\hat{\epsilon}_1$ the permittivity,
and $W_E=\int\epsilon_0\vec{E}(\vec{r})\times\hat{\epsilon}\vec{E}(\vec{r})dV$ the mode full energy. The frequency component of the temperature distribution $\widetilde{T}(\omega,\vec{r})$ can be solved in the Fourier-domain heat equation,
\begin{equation}
i\Omega\rho C\widetilde{T}+k\Delta\widetilde{T}=\widetilde{P},
\end{equation}
where $\rho$ is the density, $C$ the heat capacity, and $k$ the thermal conductivity.
The cavity shift frequency takes the form,
\begin{equation}
\frac{\delta \omega_c(\Omega)}{\omega_c}=-\frac{1}{W_E}\int\widetilde{T}(\Omega,\vec{r})\epsilon_0\sqrt{\epsilon(\vec{r})}\frac{dn}{dT}(\vec{r})|\vec{E}(\vec{r})|^2dV.
\end{equation}
In Fig.~\ref{fig:responses}, we show the cavity frequency responses of three different micro-cavities, i.e. silicon OMC, silica micro-toroid and $\textrm{Si\s{3}N\s{4}}$ ring resonator, at room temperature with an absorption power $\tilde{P}(\Omega)=1\mu W$.
The cavity frequency responses deviate from the single-pole model that we use in the main text. Multiple-poles models are typically required to fit the response~\cite{qiu_floquet_2019}.
The OMC has much larger relative frequency shift at DC compared to the other two types cavities, due to the small mode volume, which also results in high frequency poles in MHz range.
In micro-toroid, despite the large size, the thermal bandwidth can be around MHz, e.g. in Ref.~\cite{riviere_cavity_2011}.
For SiN micro-resonator, the bandwidth is typically in the kHz range, e.g. in Ref.~\cite{huang_thermorefractive_2019}. The relevant material coefficients used in the FEM simulations are shown in Table~\ref{tab:coef}.
In addition to the geometric dependence, the device surroundings sometimes can even have a more significant impact on the cavity frequency thermal response, e.g. gaseous Helium, vacuum or liquid~\cite{qiu_floquet_2019,riviere_cavity_2011}. 
\begin{table}[t]
\centering
\caption{\bf Physical properties used for the FEM simulations of the cavity frequency thermal responses.}
\begin{tabular}{|c|c|c|c|}
\hline
Physical properties
& $\text{Si}_3\text{N}_4$
& Si
& $\text{Si}\text{O}_2$
\\
\hline
Density $\rho$ (\si{kg.m^{-3}})
& 3290
& 2329
& 2203
\\
Refractive index $n_0$
& 2.00
& 3.48
& 1.50\\
Thermo-optic $\mathrm{d}n/\mathrm{d}T$ (\SI{e-5}{K^{-1}}) & \SI{2.45}{} & \SI{16.0}{}& \SI{1.29}{}\\
Thermal conductivity $k$ (\si{W.m^{-1}.K^{-1}}) & 30 & 130 & 1.38\\
Specific heat capacity $C$ (\si{J.kg^{-1}.K^{-1}}) & 800 & 700 & 703\\
\hline
\end{tabular}
  \label{tab:coef}
\end{table}

 \section{Excess Noise}\label{app:excess_noise}
Excess laser noise is known to limit sideband cooling and to introduce classical correlations in linear measurements of mechanical motion, which becomes extremely critical to deep sideband cooling,
where optical filters are typically required to reject the excess laser noise around the mechanical frequency.
External cavity diode lasers are well-known to have excess noise in the GHz range, due to the damped relaxation oscillation caused by carrier population dynamics.
In our experiments, the excess laser frequency noise spectrum density $S_{\omega\omega}(\Omega)$ at frequency of $5.2\mathrm{GHz}$ is measured below $10^5 \mathrm{rad^2}\mathrm{Hz}$.
In BHD, we choose a LO power of around $6\mathrm{mW}$.
The beating between the high reflected power ($>100\mathrm{\mu W}$) and the noise from the LO can lead to an increased noise floor by $\sim 1\%$, due to the excess noise in the LO~\cite{qiu_laser_2020,qiu_quantum_2020}.

Brillouin scattering in optical fibers is known to generate excess noises, from bulk elastic waves or surface acoustic waves~\cite{beugnot_brillouin_2014}.
In our experiments, we don't observe evident excess noise in all the optical equipments, such as optical switches, mirrors and fibers, with a quantum limited detection balanced heterodyne detection.

Fundamental temperature fluctuation can in principle result in excess thermo-refractive noise (TRN) in the cavity.
TRN has received significant attentions in the gravitational wave detection community due to the high powers and low detection bandwidth and limits the ultimate sensitivity.
Over the last years, there have been continuous efforts in the characterization of TNR in integrated devices.
The TRN is estimated to have a large impact on the on-chip Kerr squeezing due to the low detection bandwidth.
In our measurements with optomechanical spectroscopy, such effects from TRN might be negligible due to the large mechanical frequency.

The optical resonant wavelength of the silicon OMC is in the bandgap of silicon, which should in principle introduce negligible optical absorption.
Due to the ultra-small mode volume and the strong cavity field enhancement, nonlinear optical dynamics can arise in the cavity at high pumping powers, such as two-photon absorption (TPA)~\cite{barclay_nonlinear_2005a}.
The TPA in the silicon cavity can introduce excess nonlinear optical loss $\kappa\s{tpa}$ and generate free carriers.
The generated free carriers in turn lead to excess optical loss $\kappa\s{fc}$ and cavity dispersion $\Delta\s{fc}$.
The total optical cavity loss is thus given by $\kappa = \kappa\s{ex}+\kappa_s+\kappa_a+\kappa\s{tpa}+\kappa\s{fc}$,
where $\kappa\s{tpa} = \kappa^0\s{tpa} \bar{n}_c$ and $\kappa\s{fc} \simeq \kappa^0\s{fc} \bar{n}^2_c$.
At room temperature, the estimated $\kappa^0\s{tpa}/2\pi \sim 3\mathrm{kHz}$ and $\kappa^0\s{fc}/2\pi< 10\mathrm{Hz}$, taking the parameters in Ref.~\cite{sun_nonlinear_2013a,barclay_nonlinear_2005a,wu_mesoscopic_2017} and estimated mode volume from finite element simulation. We note that, $\kappa^0\s{fc}$ also depends on the carrier lifetime. In our estimation, we assume a carrier lifetime of $\sim 10\mathrm{ns}$, which in practice depends on the density of the carrier in the optical cavity~\cite{barclay_nonlinear_2005a}. At cryogenic temperatures, the carrier lifetime can decreases by few orders of magnitude compared to room temperature~\cite{sun_nonlinear_2013a}, which results in lower $\kappa^0\s{fc}$.
The optical cavity dispersion due to the free carrier is estimated $\Delta^0\s{fc}<10\mathrm{Hz}$~\cite{sun_nonlinear_2013a,barclay_nonlinear_2005a,wu_mesoscopic_2017}.
The nonlinear dynamics due to TPA and FC can in principle also result in complex dissipative dynamics in the optical cavity, which is not considered in the main text.

We note that, theoretical predictions show that the free-carriers dispersion in the optical cavity can introduce excess noise due to the incoherent carrier excitation and decay processes~\cite{hamerly_quantum_2015}, which may partially explain the noise floor we observe in the experiments and requires further investigation.

%

\end{document}